\documentstyle[nato,amsfonts]{crckapb} 
\newcommand{\bowstar}{\bowtie\kern-0.8em |~}
\def\lp{\stackrel{\leftarrow}{\partial}}
\def\rp{\stackrel{\rightarrow}{\partial}}
\def\bowstar{\bowtie\kern-0.8em |~}
\def\lb{\{ \kern-.18em | }
\def\rb{| \kern-.18em \} }
\begin{opening}
\title{A SURVEY OF STAR PRODUCT GEOMETRY}
\subtitle{NATO ARW: UIC 2000, July 22-26 \qquad \qquad ANL-HEP-CP-00-085}
\author{COSMAS ZACHOS} 
\institute{High Energy Physics Division,\\
Argonne National Laboratory, \\
Argonne, IL 60439-4815, USA \\
\phantom{.} \qquad\qquad{\sf zachos@hep.anl.gov}}      
\end{opening}
\begin{document}
\begin{abstract}
A brief pedagogical survey of the star product is provided, through
Groenewold's original construction based on the Weyl correspondence. It is
then illustrated how simple Landau orbits in a constant magnetic field,
through their Dirac Brackets, define a noncommutative structure since these
brackets exponentiate to a star product---a circumstance rarely operative 
for generic Dirac Brackets. The geometric picture of the star product based
on its Fourier representation kernel is utilized in the evaluation of chains
of star products. The intuitive appreciation of their associativity and
symmetries is thereby enhanced. This construction is compared and contrasted 
with the remarkable phase-space polygon construction of Almeida.
\end {abstract}
\section{Introduction}
The noncommutative star product of Groenewold \cite{groen} 
is the linchpin of deformation (phase-space) quantization 
\cite{moyal,bayen}. Currently, it is ubiquitous in matrix systems 
and in M-physics applications of non-commutative geometry ideas, such as in 
D-branes on a ``magnetic" B-field background \cite{natied}.
This product, connecting phase-space functions $f(x,p)$ and $g(x,p)$,
is the unique associative pseudodifferential deformation \cite{bayen}
of ordinary products:
\begin{equation}
\star \equiv ~e^{i\hbar(\stackrel{\leftarrow }{\partial }_{x}
\stackrel{\rightarrow }{\partial }_{p}-\stackrel{\leftarrow }{\partial }_{p}
\stackrel{\rightarrow }{\partial }_{x})/2}~.
\end{equation}

Since the star product  involves exponentials of
derivative operators, it may be evaluated in practice through translation 
of function arguments, 
\begin{equation}
f(x,p) \star g(x,p) = f\left( x+{i\hbar\over 2}\rp_p ,~ p-{i\hbar\over 2}\rp_x
\right )~ g(x,p).
\end{equation}
When the phase-space functions involved consist of exponentials or simple 
polynomials, such star products amount to combinations of translations and
finite-order PDEs, and allow first-principles solution of concrete deformation 
quantization problems \cite{cfz}. However, for more complicated functions,  
explicit evaluations of long strings of star products 
in this language frequently appear intractable.
(There exist well-developed numerical evaluation techniques \cite{hug},
which, however, are not reviewed here.) What to do?

There may be a way out. The more practical Fourier representation 
of this product as an integral kernel
has been utilized by Baker \cite{baker}: 
$$f\star g={1\over \hbar ^2 \pi^2}\int dp^{\prime}  
dp^{\prime\prime}  dx' dx''  ~f(x',p')~g(x'',p'') \qquad \qquad \qquad \qquad 
$$
\begin{equation} 
\times  \exp \left({-2i\over \hbar} 
\left( p(x'-x'') + p'(x''-x)+p''(x-x') \right )\right) .  \label{fourier}
\end{equation}

The cyclic determinantal expression multiplying $-2i/\hbar$ in the exponent 
is twice the area of the phase-space triangle 
$({\bf r}'',{\bf r}',{\bf r})$, where ${\bf r}\equiv (x,p)$, 

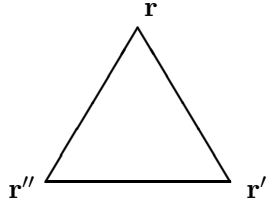
\begin{picture}(75,80)     \thicklines
\put(130,10){\line(1,0){70}}
\put(130,10){\line(3,5){35}}
\put(200,10){\line(-3,5){35}}
\put(170,75){\makebox(0,0)[cc]{{\bf r}}} 
\put(210,8){\makebox(0,0)[cc]{{\bf r}$'$}} 
\put(123,8){\makebox(0,0)[cc]{{\bf r}$''$ }} 
\end{picture}

namely, 
\begin{equation} 
2A(r'',r',r ) = ({\bf r}' -{\bf r})\wedge ({\bf r}-{\bf r}'')=
{\bf r}''\wedge {\bf r}'   +
{\bf r}'\wedge {\bf r}+
{\bf r}\wedge {\bf r}'' .
\end{equation}

For example, in this representation, it is straightforward to work out the 
distinctive hyperbolic tangent composition law of phase-space Gaussians,
$$
\exp \left (-{a \over \hbar} (x^2+p^2)\right ) ~ \star ~ 
\exp \left (-{b \over \hbar} (x^2+p^2)\right ) \qquad \qquad 
$$
\begin{equation} 
= {1\over 1+ab} 
\exp \left (-{a+b\over\hbar (1+ab)} (x^2+p^2)\right ) ,
\end{equation}
which codifies the time evolution of the harmonic oscillator \cite{bayen}.

In this representation, multiple star turn out to be simpler to 
evaluate, and the geometrical constructions they motivate exhibit conspicuously 
the symmetries and the associativity of these products. The representation 
thus rises to the level of a `picture', in Dirac's sense of a ``way 
of looking at the fundamental laws which makes their self-consistency obvious"
\cite{pamd}. Such evaluations are illustrated below, with some practical hints,
for the standard star product, as well as for some common variants and
extensions, such as the supersymmetrized version.  
This survey is based on ref \cite{triangle}, but further covers 
alternate schemes and background. 

\section{Brief Historical Review---to be skipped by experts} 

To give the general reader a flavor of how the star product is defined
in physics, some of the essentials of phase-space 
quantization which rely on it are reviewed briefly.

Weyl \cite{weyl} introduced an association rule 
mapping invertibly  c-number phase-space 
functions $f(x,p)$ (called classical kernels) to operators ${\mathfrak F}$ in 
a given ordering prescription. Specifically, 
$p\mapsto{\mathfrak p}$, $x\mapsto{\mathfrak x}$, and, in general,
\begin{equation}      
{\mathfrak F}({\mathfrak x},{\mathfrak p}) 
=\frac{1}{(2\pi)^2}\int d\tau d\sigma 
dx dp ~f(x,p) \exp (i\tau ({\mathfrak p}-p)+i\sigma ({\mathfrak x}-x)).
\end{equation}
The eponymous ordering prescription requires that an arbitrary 
operator, regarded as a power series in ${\mathfrak x}$ and ${\mathfrak p}$, 
be first 
ordered in a completely symmetrized expression in ${\mathfrak x}$ and 
${\mathfrak p}$, 
by use of Heisenberg's commutation relations, $ [{\mathfrak x},{\mathfrak p} ]
 =i\hbar $. A term with 
$m$ powers of ${\mathfrak p}$ and $n$ powers of ${\mathfrak x}$ will be 
obtained from the coefficient of $\tau^m \sigma^n$ in the expansion of 
$(\tau {\mathfrak p} + \sigma {\mathfrak x} )^{m+n}$. 
It is evident how the map yields a Weyl-ordered operator from a polynomial 
classical kernel. It includes every possible ordering with multiplicity one, 
e.g., 
\begin{equation}   
6 p^2 x^2 \mapsto {\mathfrak p}^2 {\mathfrak x}^2 + 
 {\mathfrak x}^2  {\mathfrak p}^2 + 
 {\mathfrak p}  {\mathfrak x}  {\mathfrak p}  {\mathfrak x} + 
 {\mathfrak p}  {\mathfrak x}^2  {\mathfrak p} + 
 {\mathfrak x}  {\mathfrak p}  {\mathfrak x}  {\mathfrak p} + 
 {\mathfrak x}  {\mathfrak p}^2  {\mathfrak x}~.
\end{equation}

Weyl-ordered operators clearly close among themselves under operator 
multiplication, given the degenerate Campbell-Baker-Hausdorff identity.
In a study of the uniqueness of the Schr\"{o}dinger representation,
von Neumann \cite{neumann} adumbrated the composition rule of classical 
kernels in an operator product, appreciating that Weyl's correspondence 
was in fact a homomorphism. (Effectively, he arrived at a convolution 
representation of the star product.) Finally, Groenewold 
\cite{groen} neatly worked out in detail how the classical 
kernels $f$ and $g$ of two operators ${\mathfrak F}$ and ${\mathfrak G}$ 
must compose to yield the classical kernel of ${\mathfrak F} {\mathfrak G}$,
\begin{eqnarray}  
{\mathfrak F}{\mathfrak G}&=& 
 \frac{1}{(2\pi)^4} \int {d\xi d\eta d\xi' d\eta'
dx' dx'' dp' dp''}   f(x',p') g(x'',p'')\nonumber \\
&\times  &\exp i(\xi ( {\mathfrak p}-p')+\eta ( {\mathfrak x}-x'))
\exp i(\xi' ( {\mathfrak p}-p'')+\eta' ( {\mathfrak x}-x''))  \qquad 
\end{eqnarray}  
$$
 =\frac{1}{(2\pi)^4}\!\!\int\! d\xi d\eta d\xi' d\eta' 
dx' dx'' dp' dp'' f(x',p') g(x'',p'') 
\exp i\left( (\xi +\xi')  {\mathfrak p}+(\eta+\eta')  {\mathfrak x}\right) 
$$
$$
\times ~\exp i\!\left(-\xi p'-\eta x'-\xi' 
p''-\eta'x'' +{\hbar\over 2} (\xi\eta'-\eta\xi') \right).
$$
Changing integration variables to
\begin{equation}  
\xi'\equiv {2\over \hbar} (x-x'), \quad
\xi\equiv \tau-  {2\over \hbar} (x-x'), \quad
\eta'\equiv {2\over \hbar} (p'-p), \quad
\eta\equiv \sigma- {2\over \hbar} (p'-p), 
\end{equation}
reduces the above integral to
\begin{equation}
 {\mathfrak F}{\mathfrak G}= \frac{1}{(2\pi)^2}\int d\tau d\sigma 
dx dp ~\exp i \left(\tau ( {\mathfrak p}-p)+\sigma ( {\mathfrak x}-x)\right)
~(f\star g)(x,p), 
\end{equation}
where $f\star g$ is the expression (\ref{fourier}). This is the original 
\cite{groen}, and still most physically compelling definition of the star 
product. It is this fundamental isomorphism of operator products to associative
strings of star multiplications which enables the formulation of 
Quantum Mechanics in phase space \cite{moyal,bayen}. 

\vskip 0.5cm
\noindent {\bf Remark.}~~ On a phase-space torus, 
$x+2\pi\equiv x$, $p+2\pi\equiv p$, the integer modes of periodic functions,
$f(x,p)=\exp i(m'x+n'p)$, $g(x,p)=\exp i(m''x+n''p)$, compose simply under the 
star product,
\begin{equation}
f\star g= e^{-i\hbar (m'n''-m''n')/2} ~ \exp i((m'+m'')x+(n'+n'')p).
\end{equation}
Thus, in this maximally graded basis, the antisymmetrization of the star 
product (Moyal's commutator) has a trigonometric structure constant \cite{fz}: 
$\sin \hbar(m'n''-m''n')/2 $. Its argument involves a cross product of 
two-dimensional integer-valued vectors. 
For $\hbar=4\pi/N$, this Lie algebra is identifiable \cite{fz} with $SU(N)$ 
for $N$ odd, and $SU(N/2)$ for $N$ even, and thus with $SU(\infty)$ as 
$\hbar\rightarrow 0$. 

\section{A remark on noncommutativity and Dirac Brackets } 

Parenthetically, the connection of this simple phase space to D-brane 
noncommutativity, a current application, arises as follows. 
The archetype of noncommutative geometry 
induced by magnetic flux is classical Landau orbital motion 
of a massless particle on a plane, in a constant magnetic field background.
Thus, for a vector potential $A^i= x^j \epsilon^{ji} B/2$,
and suppressing the kinetic term (which may be done consistently \cite{dunne}),
and morevoer choosing, e.g., a harmonic scalar potential \cite{dunne}:
\begin{equation}
L=\frac{B}{2} x^i \epsilon^{ij} \frac{dx^j}{dt} - \frac{k}{2} {\bf x}^2, 
\end{equation} 
whence the canonical momenta are constrained to the respective transverse 
coordinates,
\begin{equation} 
p^i=-\frac{B}{2} \epsilon^{ij} x^j.
 \end{equation}   
This amounts to two second class constraints, so the Poisson Brackets
$\{ x^i, p^j \}=\delta^{ij}$, $ \{x^i, x^j  \}=0$, and $ \{p^i, p^j  \}=0$ 
must be upgraded to Dirac Brackets, instead, for consistency:
\begin{eqnarray}
\lb  x^i , x^j \rb &=&\{x^i,x^j\} -\{x^i,~p^k +\frac{B}{2} \epsilon^{kl}x^l \}
 \frac{\epsilon^{mk}}{B} 
\{p^m +\frac{B}{2} \epsilon^{mn}x^n ,~x^j\}\nonumber \\
&=& -\frac{\epsilon ^{ij}}{B}~, 
\end{eqnarray}   
and likewise,
\begin{equation}
\lb  x^i, p^j   \rb  =\frac{\delta^{ij}}{2}, \qquad 
\lb  p^i, p^j   \rb  =-\frac{B}{4} \epsilon^{ij} ~. 
\end{equation}   
 For these particular expressions to hold, it is crucial that the 
magnetic field $B$ be a constant. 

Thus, the Hamiltonian that results from the Lagrangian  (which is 
linear in the velocity), $H=\frac{k}{2} {\bf x}^2$, yields the correct 
equations of motion describing the cyclotron orbits,
\begin{equation}
\frac {dx^i}{dt} =\lb x^i,H\rb =-\frac{k}{B} \epsilon^{ij}x^j .
\end{equation}   
The two directions $x$ and $y$, then, do not commute,
\begin{equation} 
\lb x,y \rb = -\frac{1}{B}~,
 \end{equation}   
so that perpendicular directions behave as canonical momenta to each other.

Consequently, such a plane maps to the elementary phase space used for 
illustration in this talk, and the Dirac Bracket maps to the Moyal Bracket
(the antisymmetrization of the star product already mentioned). 

The reason these Dirac Brackets exponentiate effortlessly to produce an
associative star product is because the constraints considered are linear, 
and hence these Dirac Brackets specify a Poisson manifold with 
{\em constant} bracket kernel:
$$
\lb f,g \rb = f~(\lp_{x_i} \frac{1}{2} \rp_{p_i} 
-\lp_{p_i} \frac{1}{2} \rp_{x_i} -\lp_{x_i} \frac{\epsilon^{ij}}{B} \rp_{x_j} 
-\lp_{p_i} \frac{\epsilon^{ij} B}{4} \rp_{p_j}) ~g 
$$
\begin{equation} 
\equiv \partial_i f ~J_{ij}~ \partial_j g ~. 
 \end{equation}   
In the example considered, the space where the antisymmetric matrix $J_{ij}$
acts is 4-dimensional  ($z_{2i-i}=x_i$, $z_{2i}=p_i$). 

Because $J_{ij}$ here is constant, the system  is one linear 
transformation away from a reduced standard phase space of {\em one} $x,p$ 
pair, \linebreak 
$x=x_1 + (2/B) p_2$, $p=p_1/2-(B/4)x_2$, effectively governed by Poisson
(not Dirac) Brackets, which exponentiate associatively (e.g., see the next 
section).  The exponential 
\begin{equation} 
e^{i\hbar \lp_i J_{ij} \rp_j}
 \end{equation}   
is manifestly associative and hence defines a good star product. 
This is the type of star product exploited in the current literature on 
non-commutative applications to M-theory. 
 
In sharp contrast, for {\em nonlinear}  constraints, i.e.\ {\em nonconstant} 
$J_{ij}({\bf x},{\bf p})$, exponentiation of the relevant 
Dirac Bracket kernel does {\em not} yield an associative star product,
in general. (This is illustrated for hyperspherical phase 
space in \cite{kyotozc}, with brackets 
\begin{equation} 
\lb x_i, x_j \rb =0, ~~~~\lb x_i, p_j \rb =\delta_{ij} - x_i x_j, 
~~~ \lb p_i, p_j \rb = x_j p_i- x_i p_j~, 
\end{equation} 
and whence bracket kernel 
\begin{equation} 
\lp_x\!\cdot\!\rp_p - \lp_x\!\cdot x  x\cdot\!\rp_p -\lp_p\!\cdot\rp_x 
+\lp_p\!\cdot x x\cdot \rp_x + \lp_p\!\cdot p x\cdot\!\rp_p
-\lp_p\!\cdot x p\cdot\!\rp_p.
\end{equation} 
It is at variance with the generally {\em non}-associative 
proposal of \cite{martinez}.)
That is, even though the Dirac Bracket satisfies the Jacobi identity
\cite{bracketpamd}, in this language, 
\begin{equation} 
(\partial_k J_{ij})J_{kl} 
+(\partial_k J_{jl} )J_{ki} 
+(\partial_k J_{li})J_{kj} =0, 
 \end{equation}
nevertheless, the exponential    
$~\exp\left( i\hbar \lp_i J_{ij} \rp_j\right)~$ 
itself fails associativity, in general. 

Kontsevich \cite{kontsevich} discovered, instead, elaborate
graphical rules for the generation of the appropriate 
associative star product as a series in $\hbar$.  The series starts as
$$
1+{i\hbar \lp_i J_{ij} \rp_j}
-\frac{\hbar^2}{2}\left (  \lp_i \lp_k J_{ij} J_{kl} \rp_j \rp_l \right) 
\qquad \qquad \qquad \qquad 
$$
\begin{equation} 
-\frac{\hbar^2}{3}\left (  \lp_i \lp_k J_{ij} (\partial_j J_{kl}) \rp_l 
- \lp_k J_{ij} (\partial_j J_{kl}) \rp_i \rp_l  \right) 
+O(\hbar^3).
\end{equation}   
The departure from the exponential is apparent in the second $O(\hbar^2)$ term.

\section{Composition of star products } 
In the Fourier representation, a triple star product can 
be expressed relatively simply \cite{triangle},    \pagebreak  
$$
(f\star g) \star h={1\over \hbar ^4 \pi^4}\!\int\! d\overline{p} dp' dp'' dp''' 
d\overline{x} dx' dx'' dx'''  
f(x',p') g(x'',p'') h(x''',p''') 
$$
\begin{equation} 
\times\exp {-4i\over \hbar} \left( 
A(r'', r',\overline{r})+A(r''', \overline{r}, r ) \right) .
\end{equation}
Fortunately, the intermediate $d\overline{x} ~d\overline{p}$ integrations 
collapse to mere $\delta$-functions:
$$
(f\star   g  \star h)  (x,p) =
{1\over \hbar ^2 \pi^2}\! \int \! dp' dp'' dp''' dx' dx'' dx'''  
f(x',p') g(x'',p'') h(x''',p''')
$$
\begin{equation}  
\times  \delta (x-x'+x''-x''') \delta (p-p'+p''-p''')
~\exp \left({-4i\over \hbar} A(r''', r'',r') \right)    .
\end{equation}

The product thus hinges on a triangle whose area enters in the phase of the 
exponential. The effective phase-space argument ${\bf r}=(x,p)$ of the 
product is now rigidly constrained: it lies on the new vertex of the 
parallelogram resulting from doubling 
up the triangle $({\bf r''', r'', r'})$, such that ${\bf r'-r'''}$ is one 
diagonal; the argument ${\bf r}$ lies at the end of the {\em other} diagonal, 
across ${\bf r''}$,

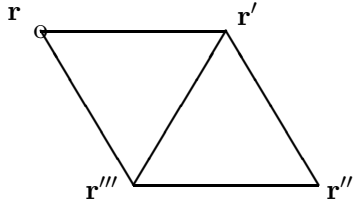
\begin{picture}(200,82)  \thicklines
\put(75,68){\line(1,0){70}}
\put(110,10){\line(1,0){70}}
\put(110,10){\line(3,5){35}}
\put(180,10){\line(-3,5){35}}
\put(110,10){\line(-3,5){35}}
\put(155,75){\makebox(0,0)[cc]{{\bf r}$'$ }} 
\put(190,9){\makebox(0,0)[cc]{{\bf r}$''$ }} 
\put(100,9){\makebox(0,0)[cc]{{\bf r}$'''$ }} 
\put(65,75){\makebox(0,0)[cc]{{\bf r}}} 
\put(75,68){\makebox(0,0)[cc]{o}} 
\end{picture}

It is then straightforward to note how this expression bears no memory of 
the grouping (order of association) in which the two $\star$-multiplications 
were performed, since the vertex  ${\bf r}$ of the parallelogram is reached 
from ${\bf r'''}$
by translating through ${\bf r'-r''}$, or, equivalently, from ${\bf r'}$
by translating through ${\bf r'''-r''}$. As a result \cite{triangle}, 
this may well realize the briefest  
graphic proof of the distinctive associativity property of the star product, 
\begin{equation} 
(f\star g) \star h= f\star (g \star h) ~.
\end{equation}

The symmetries of the triple star product, (1-3 complex conjugacy; 
cyclicity in the phase and alternating cyclicity in the effective argument;
etc.) are now evident by inspection. E.g., for $f=h$, the triple product is 
real.

Moreover, 
integration of this triple product with respect to its argument 
${\bf r}$ (tracing), e.g.\ to yield a lagrangian interaction term,  trivially 
eliminates the $\delta$-function to result in a compact cyclic expression 
of the above triangle construction for the three functions star-multiplied,
\begin{equation} 
\int \! dxdp~ f\star g \star h=
{1\over \hbar ^2 \pi^2}\! \int \! d{\bf r}_1 d{\bf r}_2 d{\bf r}_3 
f({\bf r}_1) g({\bf r}_2) h({\bf r}_3) 
~\exp\! \left({-4i\over \hbar} A(r_3 , r_2,r_1 ) \right) .  
\end{equation}

A four function star product (with three stars) involves the sum 
of the areas of two triangles, $({\bf r}_3,{\bf r}_2,{\bf r}_1$) and 
$({\bf r},{\bf r}_4,{\bf r}_1-{\bf r}_2+{\bf r}_3 )$. 
A five function star product involves the exponential of the sum of areas of 
two triangles, $({\bf r}_3,{\bf r}_2,{\bf r}_1$) and 
$({\bf r}_5 , {\bf r}_4,{\bf r}_1-{\bf r}_2+{\bf r}_3 )$, with the effective
argument restricted by 
$\delta({\bf r} -{\bf r}_1+{\bf r}_2-{\bf r}_3+{\bf r}_4-{\bf r}_5)\equiv 
\delta({\bf r} -{\bf s}_5 )$. 

Recursively, so on for even numbers of $\star$-multiplied functions,
the phase involving the sums  $A({\bf r}_3,{\bf r}_2,{\bf r}_1)+ 
A({\bf r}_5,{\bf r}_4,{\bf s}_3)+...+A({\bf r},{\bf r}_{2n},{\bf s}_{2n-1})$.
Note the cyclic symmetry, 
$ {\bf r}_1 \mapsto {\bf r}_2 \mapsto  \cdots {\bf r}_{2n} \mapsto {\bf r} 
\mapsto {\bf r}_1$.

Respectively, for odd numbers of functions,  
the phase involves sums $A({\bf r}_3,{\bf r}_2,{\bf r}_1)+
A({\bf r}_5,{\bf r}_4,{\bf s}_3)+ ... +
A({\bf r}_{2n+1},{\bf r}_{2n},{\bf s}_{2n-1})$, while the effective phase-space 
argument is restricted to 
${\bf r}={\bf s}_{2n+1} \equiv \sum_{m=1}^{2n+1} (-)^{m+1} {\bf r}_m$.
Note the cyclic symmetry in the phase, again, and the alternating cyclic 
structure in the effective phase-space argument.

As an illustration, consider phase-space points {\bf r}$_i$ arrayed in a 
regular zigzag pattern, 
(i.e.\ for the $\star$-multiplied functions getting support only on those
points on the zigzag). The arguments of the $\delta$-functions, 
{\bf s}$_{2n+1}$, then lie on a line, while the areas of the triangles
demarcated by these points increase in regular arithmetic progression
($A, 2A, 3 A, 4 A,...$): 

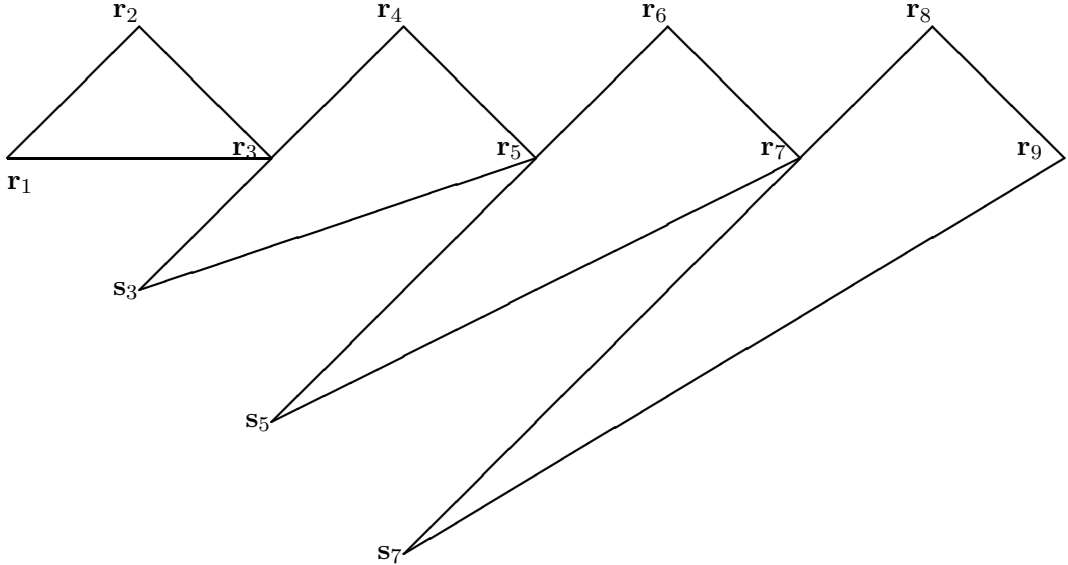
\begin{picture}(400,270)(35,0)  \thicklines
\put(0,210){\line(1,0){100}}
\put(0,210){\line(1,1){50}}
\put(50,260){\line(1,-1){50}}
\put(50,160){\line(3,1){150}}
\put(50,160){\line(1,1){100}}
\put(150,260){\line(1,-1){50}}
\put(100,110){\line(1,1){150}}
\put(100,110){\line(2,1){200}}
\put(250,260){\line(1,-1){50}}
\put(150,60){\line(1,1){200}}
\put(150,60){\line(5,3){250}}
\put(350,260){\line(1,-1){50}}
\put(5,200){\makebox(0,0)[cc]{{\bf r}$_1$}} 
\put(45,265){\makebox(0,0)[cc]{{\bf r}$_2$}} 
\put(90,213){\makebox(0,0)[cc]{{\bf r}$_3$}} 
\put(145,265){\makebox(0,0)[cc]{{\bf r}$_4$}} 
\put(190,213){\makebox(0,0)[cc]{{\bf r}$_5$}} 
\put(245,265){\makebox(0,0)[cc]{{\bf r}$_6$}} 
\put(290,213){\makebox(0,0)[cc]{{\bf r}$_7$}} 
\put(345,265){\makebox(0,0)[cc]{{\bf r}$_8$}} 
\put(387,213){\makebox(0,0)[cc]{{\bf r}$_9$}} 
\put(45,160){\makebox(0,0)[cc]{{\bf s}$_3$}}
\put(95,110){\makebox(0,0)[cc]{{\bf s}$_5$}} 
\put(145,60){\makebox(0,0)[cc]{{\bf s}$_7$}} 
\put(195,10){\makebox(0,0)[cc]{{\bf s}$_9$}} 
\end{picture}

This result is independent of the pitch of the zigzag, i.e.~the angle
at {\bf r}$_2$ ---which, in this figure, is chosen to be $\pi /2$, since this is
a local maximum of the areas $A$ of the triangles for variable pitch but
fixed lengths {\bf r}$_i -${\bf r}$_{i+1}$. One might well wonder if the 
configuration pictured could be used to define a ``classical path": its 
contribution to the phase of the exponential through the sum of all triangle 
areas, $(1+2+3+4+...) A$, is stationary with respect to 
variations such as this angle variation discussed. The question suggests 
itself, then, whether configurations stationary under {\em all} 
variations can be constructed, leading to a stationary phase evaluation 
of large/infinite star products, e.g.\ useful in evaluating 
$\star$-exponentials 
(which yield time-evolution operators in phase-space \cite{bayen}); but, so 
far, no cogent general answers appear at hand\footnote{
As an amusing curiosity, one may consider four phase-space functions,
$a({\bf r'})$, $b({\bf r''})$, $c({\bf r'''})$, $d({\bf r})$, supported 
only at the vertices of the parallelogram 
$({\bf r}, {\bf r'''}, {\bf r''}, {\bf r'})$, displayed after eqn (17). 
A product $a\star (b \star c\star a\star b\star d\star a\star c\star d)$
then repeats itself in a ``limit cycle", as additional octuple product 
factors $\star (b \star c\star a\star b\star d\star a\star c\star d)$
are appended to the right of the product, each such factor 
contributing to the phase twice the area of the parallelogram.           
}. 

\section{Almeida's Polygon} 

After completion of ref [1], an alternate, intriguing, earlier geometrical 
insight on star products was brought to my attention, \cite{almeida}, 
which organizes the problem into a different construction. 
In the conventions employed above, it
essentially  demonstrates the following \cite{almeida}:

~~~\parbox{11cm}{\sf 
For a star product  of an odd number ($2n+1$) of functions, a unique 
$2n+1$-gon is propounded in phase space, whose area equals 4 times the 
sum of the areas of triangles considered above 
($A({\bf r}_3,{\bf r}_2,{\bf r}_1)+...+A({\bf r}_{2n+1},{\bf r}_{2n},
{\bf s}_{2n-1})$) in the phase of the exponential kernel. Almeida's polygon is 
constructed as follows.

  First, {\bf s}$_{2n+1}$ (specified above)  
is constructed simply by alternating vector addition. One then extends 
the segment {\bf s}$_{2n+1}-${\bf r}$_1$ by an equal length past 
{\bf r}$_1$, to a point denoted by {\bf t}$_1$; thus, {\bf r}$_{1}$ 
lies at the midpoint of the first side of the polygon,
{\bf s}$_{2n+1}-${\bf t}$_1$. Likewise, from {\bf t}$_{1}$, one connects to 
{\bf r}$_2$, then extending to {\bf t}$_{2}$, such that {\bf r}$_2$
lies at the midpoint of the second side, {\bf t}$_{2}-${\bf t}$_1$.
So on, to {\bf t}$_{2n}$, whence one joins {\bf t}$_{2n}$ to 
{\bf s}$_{2n+1}$. It can be seen that {\bf r}$_{2n+1}$ lies on 
{\bf t}$_{2n}-${\bf s}$_{2n+1}$, and, in fact, at its midpoint. 

The polygon constructed has ${\bf r}_{1}, ...{\bf r}_{2n+1}$ at the 
midpoints of its sides.}

This polygon construction extends to an even number of 
functions $\star$-multiplied, with the effective argument serving as the 
$2n+1$th point, as before. 
For nonconvex polygons, the signed sum of convex segments
must be considered for the total area.  

This polygon is completely 
``egalitarian", in that no order of association is even apparent, which 
highlights its abstract elegance---whether it appears more 
economical in practice than the sequence of triangles considered
in the previous section, or not. 
Illustrated in the above configuration for points 
${\bf r}_{1}, ... ,{\bf r}_{5}$, it turns out to be the nonconvex pentagon\\  
(${\bf t}_4,{\bf t}_3(={\bf r}_1),~{\bf t}_2(={\bf r}_5),~{\bf t}_1,{\bf s}_5$):

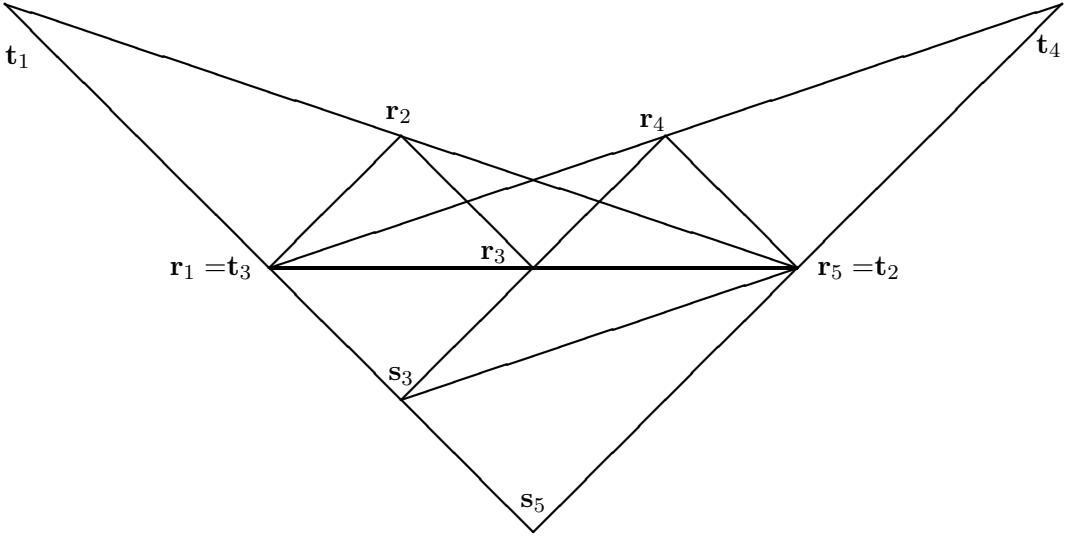
\begin{picture}(400,250)(45,0)  \thicklines
\put(100,110){\line(1,0){200}}
\put(100,110){\line(1,1){50}}
\put(150,160){\line(1,-1){50}}
\put(150,60){\line(3,1){150}}
\put(150,60){\line(1,1){100}}
\put(250,160){\line(1,-1){50}}
\put(200,10){\line(1,1){200}}
\put(0,210){\line(1,-1){200}}
\put(100,110){\line(3,1){300}}
\put(0,210){\line(3,-1){300}}
\put(78,110){\makebox(0,0)[cc]{{\bf r}$_1=${\bf t}$_3$}} 
\put(149,168){\makebox(0,0)[cc]{{\bf r}$_2$}} 
\put(185,115){\makebox(0,0)[cc]{{\bf r}$_3$}} 
\put(245,165){\makebox(0,0)[cc]{{\bf r}$_4$}} 
\put(323,110){\makebox(0,0)[cc]{{\bf r}$_5=${\bf t}$_2$}} 
\put(395,194){\makebox(0,0)[cc]{{\bf t}$_4$}} 
\put(150,69){\makebox(0,0)[cc]{{\bf s}$_3$}}
\put(200,21){\makebox(0,0)[cc]{{\bf s}$_5$}} 
\put(5,190){\makebox(0,0)[cc]{{\bf t}$_1$}} 
\end{picture}

This Almeida pentagon then has area $A({\bf t}_2,{\bf t}_1,{\bf s}_5)+
A({\bf t}_4,{\bf t}_3,{\bf t}_2)$, which, indeed, amounts to 4 times the sum of
areas of the two triangles of the recursive construction of the previous 
section, $A({\bf r}_3,{\bf r}_2,{\bf r}_1)+A({\bf s}_3,{\bf r}_5, {\bf r}_4)$. 

The simple sequence of triangles of the assemblage of the previous section 
assumes a given order of association (grouping)---but associativity has already 
been demonstrated. On the other hand, by its recursiveness, the 
addition to the existing assembly of more phase-space points, 
e.g.~{\bf r}$_6$ and {\bf r}$_7$ here, 
merely requires the evaluation of an extra triangle.
By contrast, the corresponding 
Almeida heptagon is thoroughly different, as it starts from a different 
point, {\bf s}$_7$, so that the pentagon already evaluated is not of 
particular practical significance.

\section{A Variant Product} 

A variant of the star product (cohomologically equivalent to it) 
is the lopsided associative product of Voros \cite{voros}, 
\begin{equation}
\bowstar ~\equiv ~e^{i\hbar \stackrel{\leftarrow }{\partial }_{x}
\stackrel{\rightarrow }{\partial }_{p} }~.  \label{vorosproduct}
\end{equation}

It is sometimes convenient to rotate phase-space variables canonically 
(i.e.\ preserving their Poisson Brackets),
\begin{equation}
(x,p)\mapsto(\frac{x+ip}{\sqrt{-2i}} ~, \frac{x-ip}{\sqrt{-2i}} ) ~,
\end{equation}
to represent this product as
\begin{eqnarray}
\bowstar ~&\equiv & ~e^{\hbar (\stackrel{\leftarrow }{\partial }_{x}
-i \stackrel{\leftarrow }{\partial }_{p})(\stackrel{\rightarrow }{\partial }_{x}
+i \stackrel{\rightarrow }{\partial }_{p})/2}
=e^{i\hbar (\stackrel{\leftarrow }{\partial }_{x}
\stackrel{\rightarrow }{\partial }_{p}-\stackrel{\leftarrow }{\partial }_{p}
\stackrel{\rightarrow }{\partial }_{x})/2}~
e^{i\hbar (\stackrel{\leftarrow }{\partial }_{x}
\stackrel{\rightarrow }{\partial }_{x}+\stackrel{\leftarrow }{\partial }_{p}
\stackrel{\rightarrow }{\partial }_{p})/2} \nonumber \\
&=&\star ~~e^{-\hbar (\stackrel{\leftarrow }{\partial }_{x}^2+
\stackrel{\leftarrow }{\partial }_{p}^2)/4 }~
e^{-\hbar (\stackrel{\rightarrow }{\partial }_{x}^2 +
\stackrel{\rightarrow }{\partial }_{p}^2)/4}~
e^{\hbar ((\stackrel{\leftarrow }{\partial }_{x}+
\stackrel{\rightarrow }{\partial }_{x})^2 +
(\stackrel{\leftarrow }{\partial }_{p}+
\stackrel{\rightarrow }{\partial }_{p})^2)/4}~.
\end{eqnarray}
This turns out to be the covariant transform of the star product which 
controls the dynamics when Wigner distributions are transformed into Husimi 
distributions \cite{takahashi}, 
a smoothed representation popular in applications.
  
It is plain that the Gaussian-Laplacian factors, 
\begin{equation}
T^{-1} (\partial_x,\partial_p)\equiv   
\exp(-\hbar(\partial _{x}^2 + \partial _{p}^2)/4),
\end{equation}
merely dress the standard star product into Voros' 
product \cite{voros}, 
\begin{equation}
T (f\star g)=T(f) \bowstar T(g)   ~. 
\end{equation}

The Lie algebra of brackets of $\phantom{f}\bowstar \phantom{f}$,
i.e.\ the kernel of $f\bowstar g - g\bowstar f$, starts with the 
Poisson Brackets to $O(\hbar)$. By the above equivalence, 
this Lie algebra is seen to be equivalent to the Moyal algebra \cite{moyal} 
(the algebra of brackets of $\star$, i.e.\ 
$\{\{f,g\}\}\equiv f\star g- g\star f$), 
in comportance with the general result on the essential uniqueness of the
Moyal algebra as the one-parameter deformation of the Poisson Bracket algebra
\cite{vey}.

Actually, in Fourier space, this product in its original representation 
(\ref{vorosproduct}) appears even simpler than the star product,
$$
(f\bowstar g) (x,p)={1\over 2 \pi \hbar }\int d{\bf r}^{\prime}  
d{\bf r}^{\prime\prime}  ~f(x',p')~g(x'',p'')~\delta (x''-x)
  \delta(p'-p) 
$$
\begin{equation} 
\times  \exp \left({i\over \hbar} 
(x''-x')(p'-p'') \right) .
\end{equation}
The phase-space integral is then effectively a two-dimensional $\int dx'dp''$,
 not a four-dimensional one, as the kernel has vanishing support everywhere 
but on the lines $x''=x$, $p'=p$. The triangle whose doubled area 
multiplies $-i/\hbar$ in the exponent is now a phase-space {\em right} triangle 
$({\bf r}'',{\bf r}',{\bf r})$, with its side ${\bf r}  - {\bf r'}$ 
horizontal, and its side ${\bf r}  - {\bf r''}$ vertical: 
\vskip0.3cm

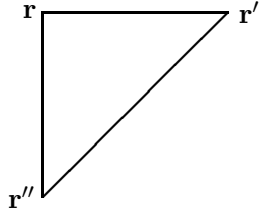
\begin{picture}(100,90)  \thicklines
\put(120,90){\line(1,0){70}}
\put(120,90){\line(0,-1){70}}
\put(120,20){\line(1,1){70}}
\put(115,90){\makebox(0,0)[cc]{{\bf r}}} 
\put(112,20){\makebox(0,0)[cc]{{\bf r}$''$}} 
\put(200,90){\makebox(0,0)[cc]{{\bf r}$'$ }} 
\end{picture}

The triple product is then seen to be actuating shifts on a rectangular 
lattice,
\begin{equation} 
(f\bowstar  g)  \bowstar h   =
{1\over (2 \pi \hbar)^2 }\!\int \! d{\bf r}^{\prime}  
d{\bf r}^{\prime\prime}  d{\bf r}^{\prime\prime \prime} 
f(x',p') g(x'',p'')h(x''',p''')~\times  \qquad 
\end{equation} 
$$
\times \delta (x'''-x) \delta(p'-p) ~\exp 
\left( {i\over \hbar} (    x'(p''-p')+ x''(p'''-p'') +x'''(p'-p''') )\right) .
$$
The phase is a cyclic expression with no memory of the order of association,
which thus proves associativity for this product, 
$(f\bowstar g)\bowstar h= f\bowstar (g\bowstar h)$. 

Pictorially, the phase is the area of the entire encompassing rectangle with 
diagonal ${\bf r}'''-{\bf s}$, minus the area of the rectangle with diagonal 
${\bf r}'-{\bf r}''$; which is also equal to the {\em  sum} of the areas of 
the rectangles with diagonals ${\bf s}'-{\bf r}'$, and ${\bf r}'''-{\bf r}''$, 
respectively. (In general, it is not twice the area of the triangle 
$({\bf r}', {\bf r}'',{\bf r}''')$.)

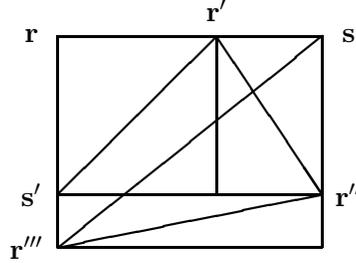
\begin{picture}(130,90)  \thicklines
\put(120,0){\line(1,0){100}}
\put(120,0){\line(0,1){80}}
\put(120,80){\line(1,0){100}}
\put(220,0){\line(0,1){80}}
\put(120,20){\line(1,0){100}} 
\put(180,20){\line(0,1){60}}
\put(120,20){\line(1,1){60}}
\put(120,0){\line(5,1){100}}
\put(120,0){\line(5,4){100}}
\put(180,80){\line(2,-3){40}}
\put(110,0){\makebox(0,0)[cc]{{\bf r}$'''$ }} 
\put(110,20){\makebox(0,0)[cc]{{\bf s}$'$}}
\put(110,80){\makebox(0,0)[cc]{{\bf r}}}
\put(180,90){\makebox(0,0)[cc]{{\bf r}$'$}}
\put(230,80){\makebox(0,0)[cc]{{\bf s}}}
\put(230,20){\makebox(0,0)[cc]{{\bf r}$''$}}
\end{picture}

\vskip 0.3cm
The construction for an n-tuple $\bowstar$-product follows simply, 
$$
{1\over (2 \pi \hbar)^n }\int d{\bf r}_1 ... d{\bf r}_n ~f_1({\bf r}_1) ...
f_n({\bf r}_n)~\delta (x_n-x) \delta(p_1-p) \qquad \qquad 
$$
\begin{equation}
\qquad \qquad \qquad \times \exp 
\left( {i\over \hbar} \sum^n_{m=1} x_m (p_{m+1}-p_m) \right) ,
\end{equation}
where $p_{n+1}$ is defined as $p_1$. A fleeting inspection of this formula
suggests an effective nearest-neighbor interaction in a natural chain. 

\vskip 0.5cm
\noindent {\bf Remark.}~~More recondite star products for particular 
nonflat phase-space manifolds, including K\"{a}hler manifolds, 
can also be formulated through integral kernels involving the Calabi function 
\cite{wakatsuki}.

\section{Graded Extension } 

~A superspace generalization of the star-product was introduced 
in ref \cite{fz}, 
(to codify the graded extension of Moyal's algebra introduced in 
ref \cite{ffz}), 
\begin{equation}
(1+ \hbar \stackrel{\leftarrow }{\partial }_{\theta}
\stackrel{\rightarrow }{\partial }_{\theta} )~\star ~ \equiv 
~\diamond ~\star~.
\end{equation}
Here, $\theta$ is the superspace Grassmann variable (nilpotent, and commuting 
with the phase-space variables): the extended star-product is then a 
direct product of the conventional piece with a superspace factor 
$1+ \hbar \stackrel{\leftarrow }{\partial }_{\theta}
\stackrel{\rightarrow }{\partial }_{\theta}$. Thus, the above extended product 
could have been alternatively written as 
\begin{equation}
e^{\hbar \stackrel{\leftarrow }{\partial }_{\theta}
\stackrel{\rightarrow }{\partial }_{\theta} }~\star ~.
\end{equation}
Hence, it can also be rewritten \cite{fradkin} 
as the evocative form, 
\begin{equation}
e^{{\frac{i\hbar }{2}}(\stackrel{\leftarrow }{\partial }_{x}
\stackrel{\rightarrow }{\partial }_{p}-\stackrel{\leftarrow }{\partial }_{p}
\stackrel{\rightarrow }{\partial }_{x}) +
\hbar \stackrel{\leftarrow }{\partial }_{\theta}
\stackrel{\rightarrow }{\partial }_{\theta}}~.
\end{equation}

Nevertheless, the original form displays associativity more readily, since 
the factor acting on the Grassmann structure is patently associative,
\begin{equation}
\left( A     \diamond B \right) \diamond 
C= A \diamond 
\left( B \diamond 
C \right),
\end{equation}   
acting on 
1d bosonic superfields $A(\theta) =a +\theta \alpha $, 
$~B(\theta) =b +\theta \beta $, so that 
\begin{equation}
A\diamond B=ab +\hbar \alpha \beta +\theta(\alpha b+ a\beta).
\end{equation}
Note the loose analogy to complex multiplication $\overline{z}_1 z_2$.
Even though this analogy cannot rise to an isomorphism, as evident from 
its noncommutativity and longer products such as the above, still, 
it turns out to be useful for actual 
evaluation of products in collecting the Grassmann even and odd terms in 
the answer. The symmetry of this product is further displayed by 
setting $\hbar=1$ and considering standard Grassmann Fourier transforms 
from bosonic to fermionic superfields,
$\tilde{A}(\theta)=\int d\phi (1+\phi\theta)~ A(\phi)
=\alpha+ \theta a $:
\begin{equation}
A\diamond B= \tilde{ A} \diamond \tilde{B}.
\end{equation}

The first of refs \cite{fradkin} provides a diagonal extension to a space of
more Grassmann variables ($N>1$ supersymmetry). 

\noindent{\bf Acknowledgments} 

Appreciation of discussions with T Curtright, D Fairlie, and  A M Ozorio de 
Almeida is recorded.
This work was supported in part by the US Department of Energy, 
Division of High Energy Physics,   Contract W-31-109-ENG-38.

\end{document}